\documentclass[amsmath,amssymb,aps,10pt,pra,nofootinbib,onecolumn,showpacs,preprint]{revtex4-1}

\usepackage{graphicx}
\usepackage{grffile} 

\usepackage{bbold}
\usepackage{color}
\definecolor{gray}{rgb}{0.5,0.5,0.5}
\usepackage[utf8]{inputenc}
\usepackage{textcomp}
\usepackage{amsmath}

\newcommand{\ksr}{\underline{K}^\mathrm{sr}}

\newcommand{\eq}[1]{Eq.~(\ref{#1})}
\newcommand{\fig}[1]{Fig.~\ref{#1}}

\graphicspath{{./figures/}}

\begin{document}
\title{\"Uberresonant Scattering of Ultracold Molecules}

\date{\today}

\author{Michael Mayle}
\author{Goulven Qu\'em\'ener}
\footnote{Present address: Laboratoire Aim\'e Cotton, Universit\'e Paris-Sud, CNRS, B\^at. 505, 91405, Orsay, France}
\author{Brandon P. Ruzic}
\author{John L. Bohn}
\affiliation{JILA, University of Colorado and National Institute of Standards and Technology, Boulder, Colorado 80309-0440, USA}

\date{\today}

\begin{abstract}
Compared to purely atomic collisions, ultracold molecular collisions
potentially support a much larger number of Fano-Feshbach resonances
due to the enormous number of ro-vibrational states available.
In fact, for alkali-metal dimers we find that the resulting density of
resonances cannot be resolved at all, even on the sub-$\mu$K temperature
scale of ultracold experiments. As a result, all observables become
averaged over many resonances and can effectively be described by
simpler, non-resonant scattering calculations. Two particular examples
are discussed: non-chemically reactive RbCs and chemically reactive
KRb. In the former case, the formation of a long-lived collision complex
may lead to the ejection of molecules from a trap.  In the latter case,
chemical reactions broaden the resonances so much that they become unobservable.
\end{abstract}

\maketitle

\section{Introduction}
The central conflict in ultracold molecular scattering is this:
On the one hand, at ultralow temperature scattering observables
are few in number, often limited to
a single two-body loss rate, sometimes complimented by an elastic cross section,
and limited to explicit information on only a small number of partial waves.
On the other hand, the underlying dynamics that drives
scattering consists of complex motion
on a three- or four- (or more-) body potential energy surface (PES).  This
surface is moreover anisotropic, so that many more angular momentum
states may contribute to scattering than the few represented by
the asymptotic partial waves.  How to properly distill the elaborate
dynamics of the collision complex into observables remains an open
question.
\footnote{In cases where the molecules are chemically reactive
at these temperatures, much more information could of course be
extracted by state-selectively detecting the products of reaction,
a task that has not yet been performed experimentally.}

For ultracold scattering of alkali-metal atoms, the link between the
PES and observables is
cemented by the observation of Fano-Feshbach resonances.  Here, the
two-body PES's involved are comparatively simple, and the remaining
undetermined parameters, consisting most simply of a pair of scattering lengths
and a $C_6$ coefficient, are used to fit data and to produce
predictive models \cite{RevModPhys.82.1225}.
Likewise, the observation of resonances should assist in interpreting
collisions of molecules with light, isotropic partners such as
helium \cite{PhysRevA.78.022705}, or perhaps even light molecules
colliding with each other \cite{Tscherbul11_NJP}.
In such a case {\it ab initio} potentials are likely to
represent something close to reality, and to be readily
fine-tuned by fitting to resonances.

In a recent paper we have begun to explore the role of resonant
scattering on heavier species with highly anisotropic interactions, specifically,
alkali-metal atoms colliding with alkali-metal dimers, at low temperature \cite{yayus}.
A main conclusion of Ref.\ \cite{yayus} was that the
density of states (DOS) of rotational and vibrational motion of the
three-atom complex may be quite high, e.g., perhaps of order
1 per Gauss in Rb + KRb scattering.  While these resonances may
conceivably be resolvable experimentally, it is likely an impossible
and unrewarding task to generate them explicitly from a PES.
Rather, Ref.\ \cite{yayus} adopted a statistical treatment of the resonances,
asking what properties of the complex could be assessed on
average.  A unifying concept in this analysis was the mean decay
width of the resonances, as given by the Rice-Ramsperger-Kassel-Marcus (RRKM)
 expression found in chemical transition state theory \cite{levinebook}
\begin{eqnarray}
\label{RRKM}
\Gamma_{\rm RRKM} = \frac{ N_o }{ 2 \pi \rho },
\end{eqnarray}
where $\rho$ is the  density of states in the vicinity of the collision
energy, and $N_o$ is the number of open scattering channels.
In ultracold collisions involving alkali-metal molecules, a large value of
$\rho$ and a small value of $N_o$ (perhaps even $N_o=1$) implies a
dense forest of very narrow resonances.

In the present paper we extend this analysis to collisions of
pairs of alkali-metal dimers.
A main finding is that the DOS for the four-atom complex is vastly
larger than for the 3-atom complex, so that the resonances so formed
cannot be resolved at all, even on the $T = 0.1 - 1$ $\mu$K temperature
scale of experiments.  In this ``\"uberresonant'' regime,
 all observables become averaged over
many resonances, effectively bypassing the inherent
intricacy of the complex.  The resulting non-resonant cross sections
are then in principle actually easier to compute and interpret than
the atom-molecule case.

We apply this idea to two cases.  One case is RbCs, which is not chemially
reactive at ultralow temperature, and for which therefore $N_o = 1$ in
its absolute, that is, ro-vibrational and spin ground state.
In this case, the resulting extremely
narrow resonances imply long complex lifetimes, potentially on the order of
experimental times. This means not only that some fraction of the
molecules remain ``invisible,'' hidden inside four-body complexes, but also that
the complexes, upon colliding with another molecule, can be ejected from
the trap, leading to an unwelcome delayed-three-body loss mechanism.
In this article we provide estimates of the loss rates implied by this
mechanism, including the effect of electric fields.
A second example is afforded by  KRb molecules,
which remain chemically reactive even at ultracold
temperature \cite{PhysRevA.82.042707}.  In this case $N_o$ includes all possible channels
of the products of reaction, and is quite large.  Thus the resonance width
implied by (\ref{RRKM}) is far larger than the mean resonance
spacing, and resonances are expected to be unobservable in the loss rates.

\section{Theoretical Model}

We consider collisions of diatomic molecules
AB (where A and B denote alkali-metal atoms) in their $^1 \Sigma$ electronic
ground state, their $v=0$ vibrational ground state, their $n=0$
rotational ground state, and some nuclear spin states
$I_AM_{A},I_BM_{B}$,  assumed to be decoupled in a magnetic field.
We pay attention to the nuclear spins in order to completely specify the
state, and to properly account for Bose/Fermi symmetrization, but
they play little other role in the theory we describe below.
Moreover, let $|LM_L \rangle$ denote the partial wave of
the incident channel, describing the relative orbital angular momentum
of the molecules.  An important quantity in the theory is then the
total angular momentum (exclusive of the nuclear spin)
${\bf J} = {\bf n}_1 + {\bf n}_2 + {\bf L}$.  Since we consider
only asymptotic states with $n_1 = n_2=0$, the value of $J$ is identical
to the partial wave $L$ in a given incident collision channel.

Introducing the shorthand notation
\begin{eqnarray}
|a \rangle = {\cal S} &&  |^1 \Sigma, v=0, n=0, I_A M_A, I_BM_B \rangle_1
\nonumber \\
\times && |^1 \Sigma, v=0, n=0, I_A M_A, I_BM_B \rangle_2  \nonumber \\
\times && |L M_L \rangle
\end{eqnarray}
(where $S$ denotes the appropriate symmetrization for bosons or fermions),
the collision cross sections can be written in terms of the scattering matrix elements $S_{a^{\prime}a}$,
\begin{align}\label{eq:sigmageneral}
\sigma_{a \rightarrow a'}
= \frac { \pi }{ k^2 } \sum_{LM_L L^{\prime}M_L^{\prime}}\left| 1-S_{a^\prime a} \right|^2\Delta.
\end{align}
$k$ is the wave number of the colliding molecules and $\Delta$ accounts for
their indistinguishability, that is, $\Delta=2$ if they are in identical
states and $\Delta=1$ otherwise. The indices $a,a'$ summarize the
quantum numbers of AB in the incident channel, and are extended to include the
product channels in the case of reactive collisions.
Even in an electric field, the projection of the total angular momentum
onto the field axis is conserved.

Following Ref \cite{yayus}, we construct a scattering
theory that incorporates
both a high density of resonant states of the collision complex, and
threshold effects relevant to ultralow energies.
This is achieved by combining multichannel quantum defect theory (MQDT) with
the methods of random matrix theory. In doing so, we exploit the
conceptual difference between the spin channels $|a \rangle$ that
describe physics at large interparticle separation $R$; and the numerous
resonant states of the complex, denoted $|\mu \rangle$, that differ by
rotational and vibrational quantum numbers from $a$. The key feature
of MQDT is that one only needs to provide the reactance matrix
$\ksr$ which is defined at
a ``matching radius'' $R_m$ that defines the boundary between short- and
long-range physics.
The MQDT formalism as outlined in Refs.\
\cite{PhysRevLett.81.3355,yayus,ruzicunpublished}
accounts exactly for the wave functions for
$R>R_m$ and directly yields the physical scattering matrix
$\underline S^\text{phys}$ via standard algebraic procedures.

As in our previous work, for $R>R_m$ we assume simplified long-range
interactions of the form
\begin{equation}
V_a(R) = - \frac{ C_6 }{ R^6 } + \frac{ \hbar^2 L_a(L_a+1) } {2 m_r R^2}+E_a,
\end{equation}
where $E_a$ is the threshold of the $a$th channel, which may depend
on a magnetic field $B$. Here, $m_r$ is the reduced mass of the
scattering partners and $C_6$ is their van der Waals coefficient,
which is taken to be isotropic in this model.  These potentials
are used to calculate the relevant MQDT parameters from which
the cross sections are ultimately constructed.  We will see below how to account
for nonzero electric fields.

The short-range $K$-matrix is constructed according to the dictates
of random matrix theory \cite{RevModPhys.82.2845} as
\begin{equation}\label{eq:ksr}
 K^\text{sr}_{a'a}(E)=-\pi\sum_{\mu=1}^N\frac{W_{a'\mu}W_{\mu a}}{E-E_\mu}.
\end{equation}
It is indexed by the $N_a$ asymptotic channels $a$, but is influenced
by the myriad (i.e., $N\gg N_a$) resonant states $\mu$.
The input parameters for the resonant scattering theory,
\eq{eq:ksr}, are the zero-order positions $E_\mu$ of the resonances
and the coupling elements $W_{a\mu}$ to the asymptotic channels.
Within our statistical framework, $E_\mu$ and $W_{a\mu}$ are taken as
random variables based on the Gaussian Orthogonal Ensemble (GOE)
\cite{yayus,RevModPhys.82.2845}. By employing such a
model, we assume that the collision complex corresponds classically
to a long, chaotic trajectory that ergodically explores a large portion
of the allowed phase space.

The GOE is in turn specified by the mean resonance width.
It was determined in \cite{yayus} that a reasonable
approximation for this width is the RRKM result itself,
\begin{equation}\label{eq:kRRKM}
 \Gamma=\frac{N_a}{2\pi\rho}=\Gamma_\mathrm{RRKM},
\end{equation}
where $N_a$ is the total number of asymptotic channels in the $v=0$, $n=0$
ground state manifold.  Further narrowing of the resonances due to the
Wigner threshold laws is accounted for within the MQDT theory.

Thus the resonance model is completely specified by the density of
states $\rho$.  We estimate the DOS in the same way as in Ref.\ \cite{yayus}.
Namely, we posit a set of approximate potential curves
\begin{eqnarray}
\label{LJ}
V_\mathrm{LJ}(R) + \frac{ \hbar^2 L_c(L_c+1) }{ 2m_{r} R^2 } +
E_\mathrm{rv}(v_{c1},n_{c1},v_{c2},n_{c2}).
\end{eqnarray}
Here $V_\mathrm{LJ}(R)$ is a Lennard-Jones potential with the correct $C_6$ for the
molecule-molecule interaction, and tuned to a depth equal to the binding
of the A$_2$B$_2$ complex relative to the AB + AB threshold.  This potential
is augmented by a partial wave $L_c$  of the complex, and by
a threshold energy $E_\mathrm{rv}$
corresponding to ro-vibrational excited states of the molecules in
the complex.  Key to our DOS approximation is that {\it all}
possible states that preserve the total angular momentum
${\bf J} = {\bf n}_{c1} + {\bf n}_{c2} + {\bf L}_c$ and conserve energy
are included.
Although the total $J$ is limited to a few values as dictated by the
incident partial wave $L$ of scattering, the angular momentum
quantum numbers of the complex can span into the hundreds (see below).

Having identified all such relevant potentials (\ref{LJ}),
we compute their bound states lying
near the incident threshold, and by counting them determine the DOS.
The complete DOS thus constructed assumes ergodicity, i.e., that all
states not forbidden by conservation laws are actually potentially populated.
However, this assumption can be adjusted by, say, reducing the maximum value of
orbital angular momentum $L_c$ used in the estimate.

\section{Four-body density of states}

\begin{table}
\caption{Ro-vibrational densities of states for $M=0$ and corresponding one-open-channel RRKM lifetime $\tau_0=2\pi\hbar\rho$; for $N_o>1$ one has $\tau=\tau_0/N_o$. Regarding the magnetic dipole moment, in the case of (fermionic) KRb we assumed for odd $J$ that the both molecules are in their lowest state ($M_\mathrm{K}=-4$, $M_\mathrm{Rb}=3/2$); for even $J$ one molecule is considered to be in this lowest state, the other in the next higher one ($M_\mathrm{K}=-3$, $M_\mathrm{Rb}=3/2$). For (bosonic) RbCs it is the other way round: for even $J$ they are in the same state ($M_\mathrm{Rb}=3/2$, $M_\mathrm{Cs}=5/2$) and for odd $J$ in the next higher one ($M_\mathrm{Rb}=3/2$, $M_\mathrm{Cs}=7/2$). Values used for the magnetic moments are \cite{aldegunde:033434}:
($^{40}$K$^{87}$Rb) $M_\mathrm{K}=-4$, $M_\mathrm{Rb}=3/2$, $\mu_\mathrm{mag}=2.84$\,kHz/G;
$M_\mathrm{K}=-3$, $M_\mathrm{Rb}=3/2$, $\mu_\mathrm{mag}=3.08$\,kHz/G;
($^{87}$Rb$^{133}$Cs) $M_\mathrm{Rb}=3/2$, $M_\mathrm{Cs}=5/2$, $\mu_\mathrm{mag}=3.50$\,kHz/G;
$M_\mathrm{Rb}=3/2$, $M_\mathrm{Cs}=7/2$, $\mu_\mathrm{mag}=4.07$\,kHz/G.
\label{tab:dos}
}
\begin{ruledtabular}
\begin{tabular}{lcccc}
molecule&$J$&$\rho(\mu\mathrm{K}^{-1})$&$\rho(G^{-1})$&$\tau_0 (\mathrm{ms})$\\
\hline
KRb + KRb   &  0  & 3243  & 922  & 156\\
            &  1  & 9697  & 2871 & 465\\
            &  2  & 16120 & 4582 & 774\\
            &  3  & 22512 & 6666 & 1080\\
RbCs + RbCs &  0  & 942   & 368  & 45\\
            &  1  & 2812  & 1021  & 135\\
            &  2  & 4672  & 1823  & 224\\
            &  3  & 6521  & 2369 & 313\\
\end{tabular}
\end{ruledtabular}
\end{table}

We estimate the DOS as described above and in Ref.\ \cite{yayus},
for two prototypical ultracold molecules: RbCs \cite{Danzl2010}
and KRb \cite{Ni2008a,PhysRevLett.105.203001}.
To construct the Lennard-Jones potentials in Eq.~(\ref{LJ}) for these species,
we use the $C_6$ coefficients from Ref.\ \cite{1367-2630-12-7-073041},
and potential depths of 800\,cm$^{-1}$ for (RbCs)$_2$
 \cite{PhysRevA.78.022705} and 2779.6\,cm$^{-1}$ for (KRb)$_2$
\cite{PhysRevA.82.010502}.
To compute the ro-vibrational spectrum
$E_\mathrm{rv}$ we employ the empirical potential of Ref.\ \cite{PhysRevA.83.052519}
for RbCs, and that of Ref.\ \cite{pashov:022511} for KRb.

\begin{figure}
\includegraphics[width=8.5cm]{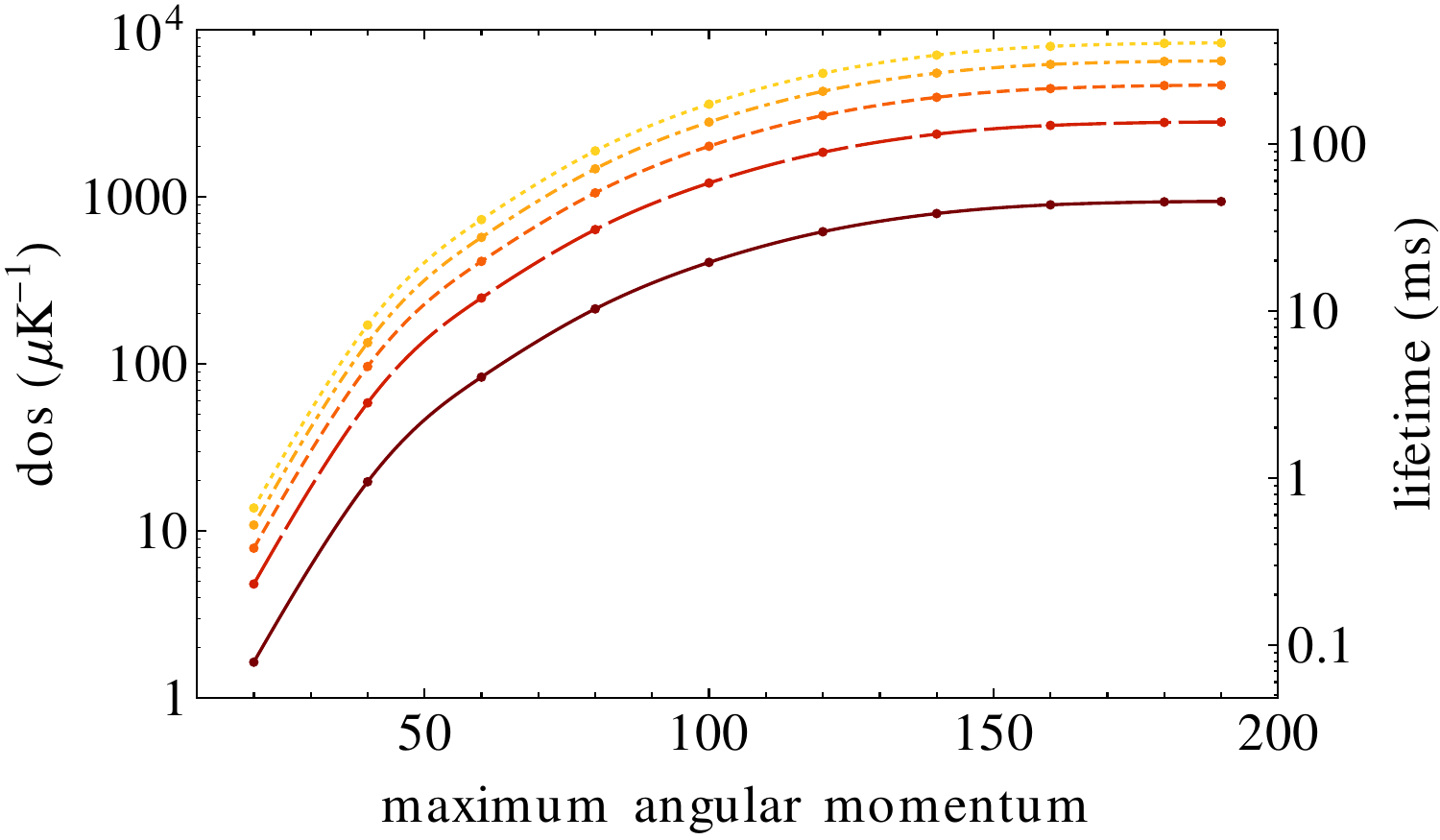}
\caption{(Color online) Ro-vibrational density of states of RbCs + RbCs collisions as a function of the maximally allowed end-over-end angular momentum $L_c$ of the collision complex, for a total angular momentum $J=0-4,M=0$ (solid, long-dashed, dashed, dashed-dotted, and dotted line, respectively). The scale on the right axis gives the corresponding RRKM lifetime of the collision complex according to $\tau_0=2\pi\hbar\rho$.
\label{fig:dos_rbcs}}
\end{figure}

The resulting ro-vibrational  DOS for several total angular momenta $J$
is reported in Table \ref{tab:dos}.
In Ref.\ \cite{yayus} the possibility for processes that change the nuclear spin
were considered, but we do not do so here; thus the table counts
only the ro-vibrational density of states.
Also shown is the mean lifetime
of the collision complex, estimated as $\tau_0 = 2 \pi \hbar \rho$.
These estimates assume that all
states of allowed angular momentum defining the complex, $L_c$, $n_{c1}$,
$n_{c2}$ in the vicinity of threshold
can actually be populated.  Even relaxing this assumption and reducing $L_c$,
the DOS remains quite high,  as seen in \fig{fig:dos_rbcs}.

\begin{figure}
\includegraphics[width=8.5cm]{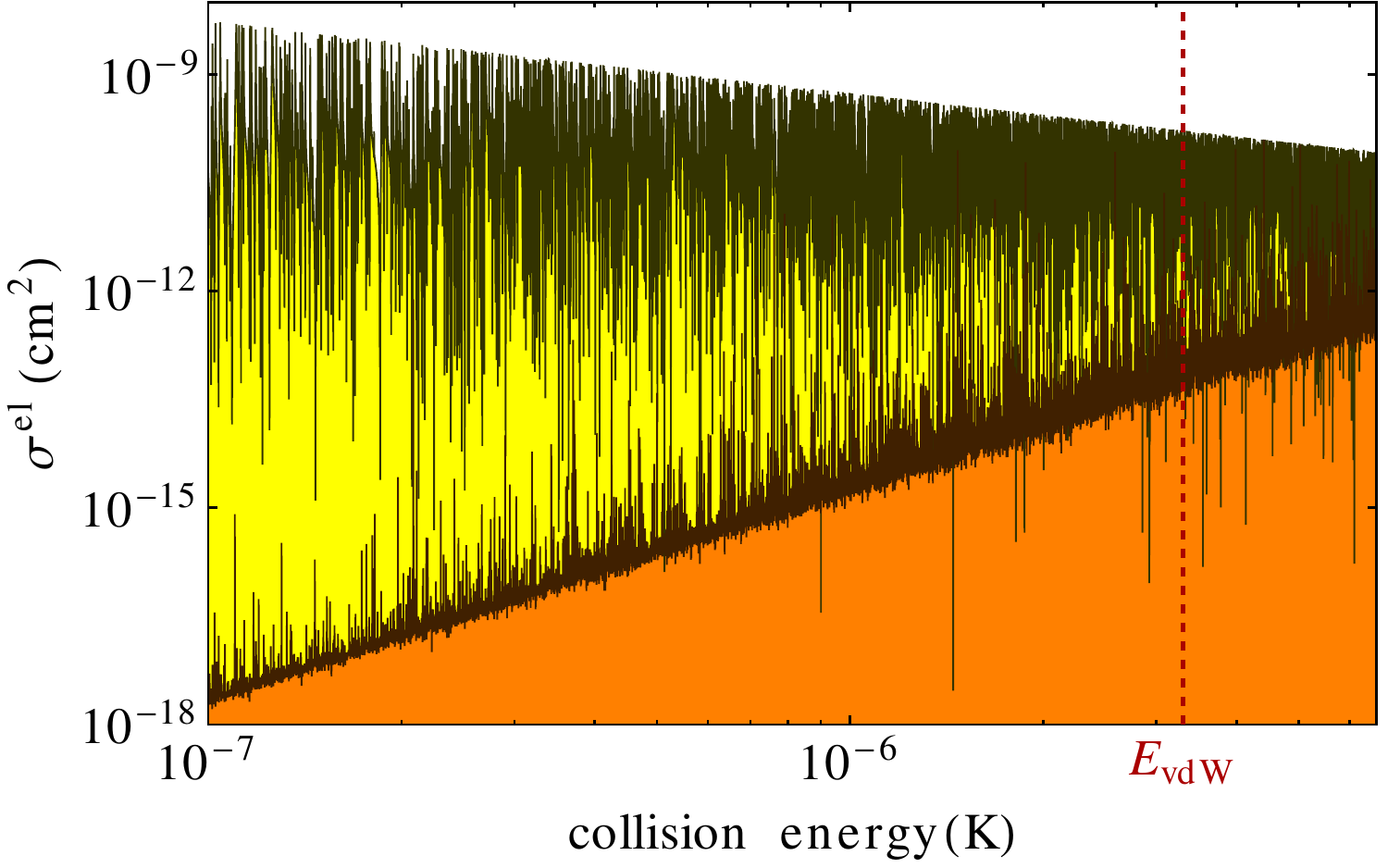}
\caption{(Color online) Elastic $s$-wave [yellow (light gray)] and $d$-wave
[orange (gray)] cross section of RbCs + RbCs. Incident channel is the absolute ground state ($v=n=0$, $M_\mathrm{Cs}=7/2$, $M_\mathrm{Rb}=3/2$ for both molecules).
\label{fig:csrbcs}}
\end{figure}

An exemplary elastic cross section for RbCs molecular collisions in the
absolute ground state ($v=n=0$, $M_\mathrm{Cs}=7/2$, $M_\mathrm{Rb}=3/2$
for both molecules) is presented in \fig{fig:csrbcs}.
Shown are cross sections for $s$-wave [yellow (light gray)]
and $d$-wave [orange (gray)]
scattering.  This figure covers an energy range of twice the
van der Waals energy $E_\mathrm{vdW}=\hbar^3 (2m_r)^{-3/2}C_6^{-1/2}$]
and contains 10,000 energy points, evenly spaced on a logarithmic grid.
At this resolution, most of the $s$-wave resonances are resolved, and the cross
section frequently approaches the unitarity limit, $8 \pi/k^2$. For
$d$-wave scattering, these resonances are not resolved so well.  It is clear from
the figure that at typical ultracold temperatures $T = 0.1 - 1$ K,
these resonances can never be resolved.

\section{Influence on scattering of non-reactive molecules}

The extremely high density of states estimated in the previous section implies
a striking feature of molecule-molecule cold collisions.  Namely, molecules
that meet on resonance may become lost in the complex for times on the order of
many milliseconds, comparable to the time scales of a typical experiment.
In this section we formulate a set of rate equations accounting for this
occurrence, using RbCs as an example.  The rate equations describe three
separate events: i) a pair of RbCs molecules meet and stick together,
thus temporarily transforming into four-body complexes, with
number density $n_c$; ii) The complexes decay back into
molecules on a time scale
set by the mean lifetime of the resonant states; and iii) during the lifetime
of the complex, another RbCs molecule can collide with it, leading almost
certainly to trap loss.  We deal with each of the three parts of this process
in the following.

\subsection{Molecule-sticking rate}

For RbCs in its absolute ground state, the number of open channels is
exactly $N_o=1$.  This circumstance automatically places resonant
scattering in the limit where, on average, resonance widths are
smaller than the mean resonance spacing, and resonances do not overlap.
Thus only some fraction of the collision events lead to long-lived
resonances, albeit very long-lived ones.  We model the sticking
process by ascribing to it a cross section which is zero away from
resonance, but which contains resonances at the appropriate DOS
and width distribution.  Such a cross section is in fact afforded by
the elastic cross section $\sigma^{\rm el}(E)$, which is easily
computed from our statistical MQDT formalism.

The rate at which the complex-forming collisions happen is given by
a thermally averaged rate constant, here distinguished by the
partial wave $L, M_L$ considered:
\begin{equation}\label{eq:Kthermal}
K_{\rm mm}^{(L,M_L)}(T)=\int_0^\infty \sigma_{L,M_L}^\mathrm{el}(E)\, v f(v)dv,
\end{equation}
where ``mm'' stands for ``molecule-molecule,'' and
\begin{equation}
f(v)=\sqrt{\frac{2}{\pi}}\left(\frac{k_bT}{m_r}\right)^{-\frac{3}{2}}v^2e^{-m_rv^2/2k_bT}
\end{equation}
is the Maxwell-Boltzmann distribution for the relative velocity for
a given temperature $T$ of the initial molecular sample.

When the mean resonance spacing is far less than the temperature, as
we assume, then these many resonances are averaged over.  We can therefore
replace the strongly-varying cross section by its mean value, taken
over each of the isolated resonances separately, and averaged over
the mean spacing $d = \rho^{-1}$ between resonances:
\begin{equation}
\label{eq:average}
\bar\sigma_{L,M_L}^\mathrm{el}(E)= \frac{ 1 }{ d }
\int_{E-d/2}^{E+d/2}\sigma_{L,M_L}^\mathrm{el}(\varepsilon)d\varepsilon.
\end{equation}
This amounts to saying that only a fraction of collision energies,
approximately $\Gamma(k)/d$, are on resonance and can lead to large
sticking times, where $\Gamma(k)$
is the mean resonance width in the vicinity of energy $E=\hbar^2 k^2/ 2 m_r$.
Note that the resonant cross sections scale as the unitarity limit,
$\propto 1/k^2$, whereas resonance widths $\Gamma(k) \propto
k^{2L+1}$,  leading to a threshold law $K_{\rm mm}^{L,M_L} \propto k^{2L}$
for the sticking rate.

More quantitatively, we make use of the simple algebraic structure of the
MQDT formalism, in the ultracold limit and for a single channel ($N_a=N_o=1$) the elastic cross section reads
\begin{align}
\sigma_{L,M_L}^\mathrm{el}(E)=\Delta\frac{4\pi}{2m_rE}\frac{A(E,L)^2}{[K^\mathrm{sr}(E)]^{-2}+A(E,L)^2},\label{eq:sigmaonech}
\end{align}
where $K^\mathrm{sr}=-\pi\sum_\mu W_\mu^2/(E-E_\mu)$. In the ultracold limit, the energy dependent MQDT parameter $A(E,L)$ can be written down explicitly \cite{ruzicunpublished},
\begin{equation}
A(E,L)^{1/2}=-\frac{\pi2^{-2L-3/2}}{\Gamma(\frac{L}{2}+\frac{5}{4})\Gamma(L+\frac{1}{2})}R_\mathrm{vdW}^{L+\frac{1}{2}}\left(\frac{\hbar}{\sqrt{2m_r E}}\right)^{L+\frac{1}{2}}.
\end{equation}
$R_\mathrm{vdW} = (2 m_r C_6 / \hbar^2)^{1/4}$ is the van der Waals length scale. In deriving \eq{eq:sigmaonech} we employed the ultracold limits $\mathcal{G}\rightarrow0$ and $\eta\rightarrow0$ of the remaining MQDT parameters \cite{ruzicunpublished}. In the vicinity of a resonance at $E_0$ and replacing the short- to long-range couplings $W_\mu$ by their average, $\pi W_\mu^2=\bar\Gamma/2$ \cite{yayus}, \eq{eq:sigmaonech} becomes
\begin{align}
\sigma_{L,M_L}^\mathrm{el}(E)\approx\Delta\frac{4\pi}{2m_rE}\frac{[A(E,L)\bar{\Gamma}/2]^2}{(E-E_0)^2+[A(E,L)\bar{\Gamma}/2]^2}.\label{eq:sigmasingle}
\end{align}
Assuming additionally that $A(E,L)$ is approximately constant within
the range of a single resonance, Eqs. (\ref{eq:average},\ref{eq:sigmasingle})
yield the mean cross section at  collision energy $E$,
\begin{equation}\label{eq:sigmamean}
\bar\sigma_{L,M_L}^\mathrm{el}(E)=\Delta\frac{4\pi}{2m_rE}A(E,L).
\end{equation}
We therefore identify the rate constant for collisional sticking as
\begin{align}
\bar K_{\rm mm}^{(L,M_L)}(T)&=\int_0^\infty \bar\sigma_{L,M_L}^\mathrm{el}\!(E)\, v f(v)dv\\
&= \Delta\frac{2^{-3L+2}\pi^{5/2}R_\mathrm{vdW}^{2L+1}}{\Gamma(\frac{L}{2}+\frac{1}{4})^2\Gamma(L+\frac{3}{2})}\frac{m_r^{L-1}}{\hbar^{2L-1}}(k_BT)^L.\label{eq:Kqdt}
\end{align}

Interestingly, this expression agrees {\em exactly} with the inelastic
rate constant derived for scattering in the presence of rapid loss due
to chemical reactions \cite{PhysRevA.84.062703}, modeled by assuming
unit loss probability at each collision energy.
The effect of averaging over a very large number of very narrow resonances
has produced a cross section that is equivalent to full absorption at
every collision energy, modified by the appropriate threshold laws.
This is a tremendous simplification: rather than even attempt to
deal explicitly with real potential energy surfaces and the many resonances
they engender, we are able to cut immediately to the observable
consequence, namely, temperature-dependent sticking probabilities.

\begin{figure}
\includegraphics[width=8.5cm]{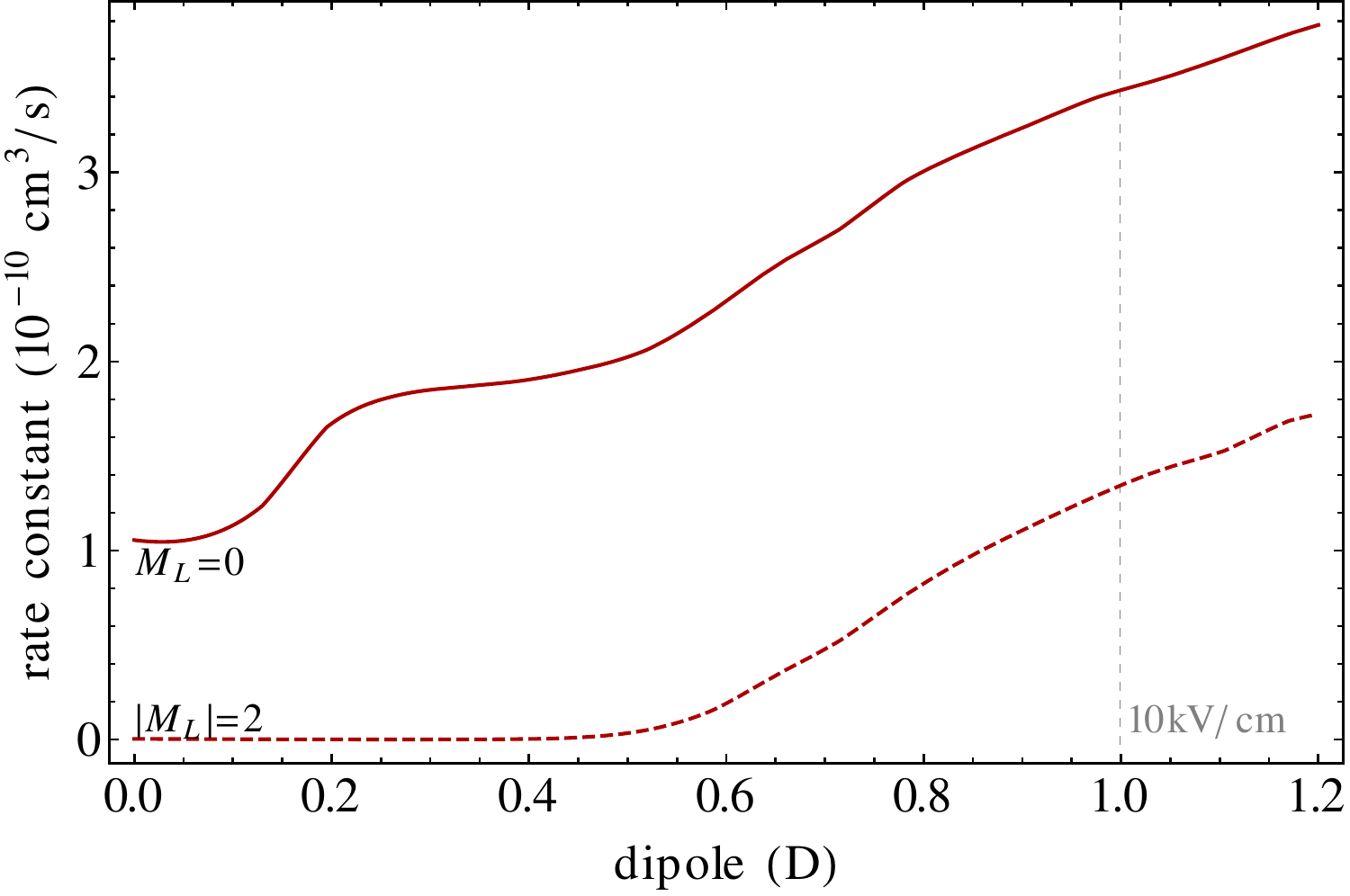}
\caption{Quenching rate constants of two indistinguishable bosonic polar $^{87}$Rb$^{133}$Cs molecules as a function of the induced dipole moment for a temperature of $T=1\,\mu$K. The solid line represents the $M_L=0$ contribution, the dashed line the $M_L=2$ (equal to the $M_L=-2$) contribution.
\label{fig:rate}}
\end{figure}

Armed with this insight, we can immediately extend the model to
nonzero electric fields, assuming that the field significantly affects
only long-range physics, and molecules reaching small $R$ vanish
with unit probability.  This problem can be solved exactly as in Refs.
 \cite{PhysRevA.81.022702,PhysRevA.84.062703}.
The resulting rate constants for our example of RbCs collisions are
reproduced in \fig{fig:rate}. References
 \cite{PhysRevA.81.022702,PhysRevA.84.062703}
predict that rate constants for loss in partial wave $L$ scales
as $d^{4L+2}$ for induced dipole moment $d$.  Hence, for small dipole
moments $s$-wave scattering
prevails. In \fig{fig:rate} the rate constant in the upper curve
shows an initial rise $\propto d^2$ for $s$-waves, until it saturates
at around $d \sim 0.25$D  \cite{PhysRevA.82.020703}.
There is then a second rise, owing to the rapid increase of
loss rate in the $d$-wave channel, which dominates the loss
beyond $d \sim 0.6$D.  For scattering with orbital angular momentum
component $M_L=2$ (lower curve), the $d$-wave rise is still apparent,
but there is, of course, no $s$-wave contribution at smaller dipole moment.

\subsection{Mean lifetime of the complex}

The resonant complexes formed in molecule-molecule collisions
will eventually decay back into
pairs of molecules.  The lifetime of the complex at a given collision
energy can be quantified by means of the time delay
\cite{Wigner55_PR,Smith60_PR,fano86},
\begin{equation}\label{eq:delay}
\tau_\mathrm{delay}=2\hbar\frac{d\delta}{dE}.
\end{equation}
Here $\delta$ is the eigenphase sum, that is, the sum of the inverse
tangents of the eigenvalues of the $K$-matrix. Employing the same
approximations as in deriving Eqs.~(\ref{eq:sigmasingle},\ref{eq:sigmamean}),
the time delay close to a resonance at $E_0$ reads
\begin{equation}\label{eq:delaysingel}
\tau_\mathrm{delay}\approx2\hbar\frac{A(E,L)\bar\Gamma/2}{(E-E_0)^2+[A(E,L)\bar\Gamma/2]^2},
\end{equation}
and therefore the mean time delay becomes
\begin{equation}
\bar\tau_\mathrm{delay}=\rho\int_{E_0-1/(2\rho)}^{E_0+1/(2\rho)}2
\hbar\frac{d\delta}{dE}dE=2\pi\hbar\rho.
\end{equation}
This is just the lifetime of the resonant complex as predicted by the
RRKM theory, \eq{eq:kRRKM}.
Just as we need not consider individual resonances in
the high-density limit, neither do we need to consider their individual
lifetimes -- another simplification.  We therefore define,
for each partial wave $L$,
a decay rate of the complexes, $\gamma_L=\bar\tau_\mathrm{delay}^{-1}$, which
follows immediately from the DOS in Table \ref{tab:dos}.

\subsection{Rate equations}

We are now in a position to formulate the rate equations for the
ultracold gas of RbCs molecules.  Denote by $n_m$ the number density of these
molecules and by $n_{c,L}$ the number density of the transient four-body complexes
formed from initial partial wave $L$ of the molecule-molecule scattering.
Because the molecule-molecule scattering rates $K_{\rm mm}^{(L,M_L)}$ and the decay rates
are different for different $L$, we explicitly add together the different
contributions, as if they were independent.
We assume that molecule-complex collisions are $s$-wave dominated and
field independent, and hence described by a universal rate of the form
\eq{eq:Kqdt}, with $L=0$ and appropriate values
for the reduced mass  and $C_6$.
Rate equations that describe the sticking of two molecules to form
the complex, the subsequent decay of the complex, and demolition of
a complex due to collision with another molecule, are given by
\begin{align}
\dot{n}_m&=\sum_{L} \left(-n_m^2\sum_{M_L}K_{\rm mm}^{(L,M_L)}+2\gamma_L n_{c,L}-K_{\rm mc}n_mn_{c,L}\right),\label{eq:nm}\\
\dot{n}_{c,L}&=\tfrac{1}{2}n_m^2\sum_{M_L}K_{\rm mm}^{(L,M_L)}-\gamma_L n_{c,L}-K_{\rm mc}n_mn_{c,L}.\label{eq:nc}
\end{align}
Here $K_{\rm mm}^{(L,M_L)}$ and $K_{\rm mc}$ are the molecule-molecule and
molecule-complex collision rate constants. As shown in Table \ref{tab:dos}, different total angular momenta $J$ lead to different densities of states and therefore different lifetimes of the collision complex (recall that $J$ is identical with the incident partial wave $L$ of the collision). This is accounted for in \eq{eq:nm} by allowing the formation of different, independent collision complexes with densities $n_{c,L}$, each possessing its own decay rate $\gamma_L$. The complexes are populated according to the molecular collision rate $K_{\rm mm}^{(L,M_L)}$ for the given partial  wave $L$ as extracted from \fig{fig:rate}. The molecule-complex collision rate constants are considered equal for all complexes. Moreover, we assume that different $M_L$ give rise to the same DOS and therefore to the same lifetime $\gamma_L$.

\begin{figure}
\includegraphics[width=8.5cm]{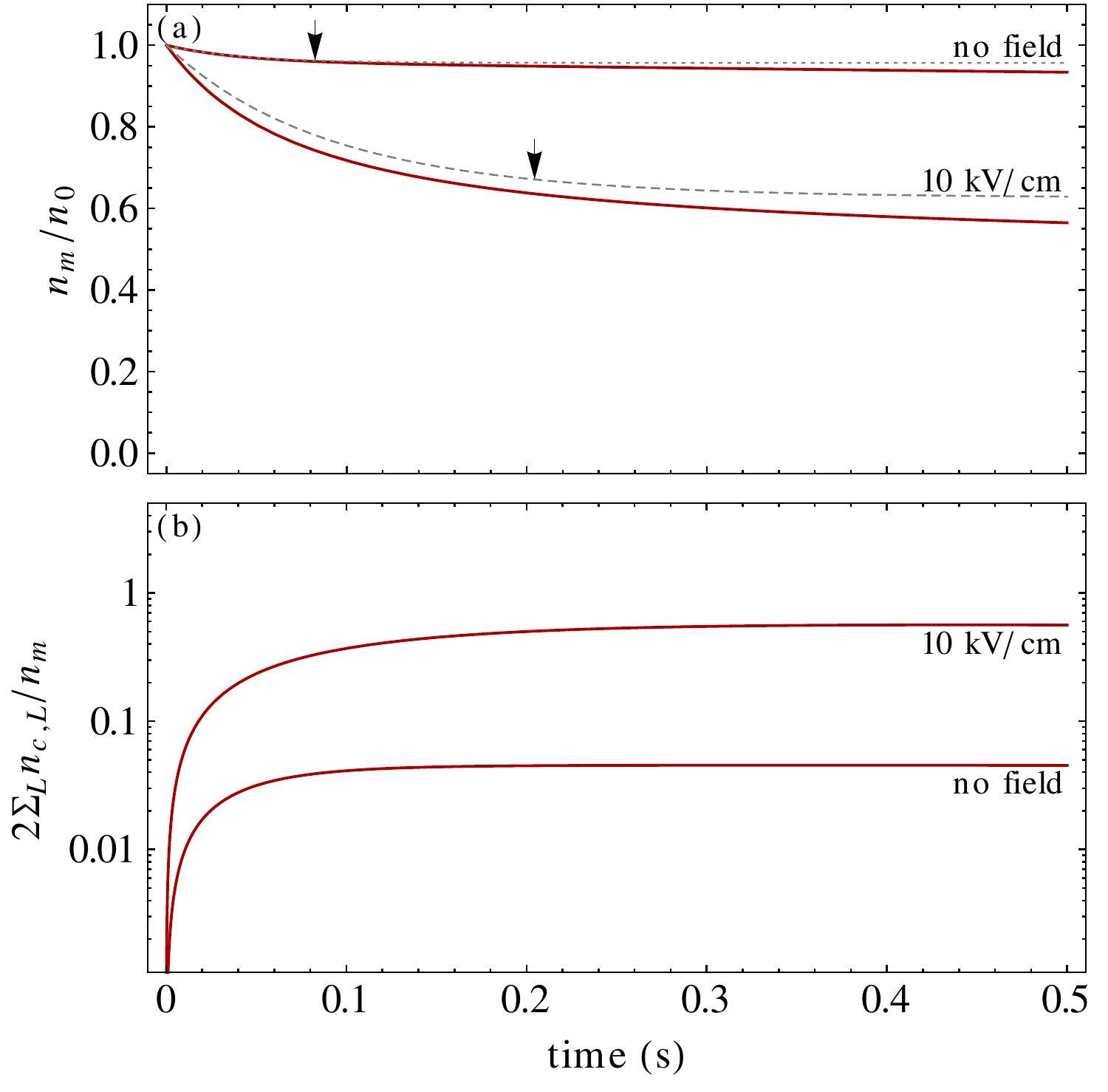}
\caption{
(a) Time evolution of the RbCs molecular density $n_m(t)$ for zero field (upper curves) and an applied electric field of 10\,kV/cm (lower curves); $s$- and $d$-wave wave collisions are considered. Dotted ($s$-wave) and dashed ($d$-wave) lines are computed without particle loss due to complex-molecule collisions, that is, $K_{\rm mc}=0$ in Eqs.\ (\ref{eq:nm},\ref{eq:nc}). The initial molecular density is $n_0=10^{10}\mathrm{cm}^{-3}$ at a assumed sample temperature of 1\,$\mu$K. The arrows indicate the timescale $t'$ as given by \eq{eq:tprime}.
(b) Evolution of the ratio of molecules bound in complexes to free molecules. The same set of parameters as in panel (a) is used.
\label{fig:moleculardensity}}
\end{figure}

The time-dependent molecular density $n_m(t)$ resulting from a numerical
integration of Eqs.\ (\ref{eq:nm},\ref{eq:nc}), starting from an initial
molecular density $n_0$,  is presented in
\fig{fig:moleculardensity}(a).  Results are shown for two different
electric field strengths
(0 and 10 kV/cm, respectively). Initially, when $n_c(t)\ll n_m(t)$, \eq{eq:nm} is dominated by the loss due to complex formation at a rate  $-K_{\rm mm}n_0^2$, where $K_{\rm mm}=\sum_{L,M_L}K_{\rm mm}^{(L,M_L)}$ is the total molecular loss rate. After some time $t'$ this initial, fast decay turns over into a slow decay due to lossy molecule-complex collisions. Some insights can be gained by setting $K_{\rm mc}=0$ for the moment, that is, no lossy molecule-complex collisions. The resulting molecular density $n_m^{(0)}(t)$ is shown in \fig{fig:moleculardensity}(a) as dotted (assuming only $s$-wave collisions) and dashed ($d$-wave collisions) lines. In this case, the solution $n_m^{(0)}(t)$ reaches an equilibrium,
\begin{align}
n_m^{(0)}(t)&\rightarrow \frac{\sqrt{1+4 n_0 K_{\rm mm}\gamma^{-1}}-1 }{2 K_{\rm mm}\gamma^{-1}},\label{eq:nmfinal}\\
\frac{n_m^{(0)}(t)^2}{n_c^{(0)}(t)}&\rightarrow\frac{2\gamma}{K_{\rm mm}},\label{eq:nmovernc}
\end{align}
as $t\gg t'$. The timescale $t'$, on which the initial linear decay turns to reach this dynamical equilibrium, can be extracted from the analytic solution $n_m^{(0)}(t)$ as
\begin{equation}\label{eq:tprime}
t'=\frac{ 2 }{ \sqrt{ \gamma^2 + K_{\rm mm} \gamma n_0} }.
\end{equation}
This time is indicated in \fig{fig:moleculardensity}(a) by arrows, assuming purely $s$-wave collisions for zero field and $d$-wave collisions for 10\,kV/cm. Moreover, by inserting \eq{eq:nmovernc} into \eq{eq:nm} one finds an expression for the slow final decay,
\begin{equation}\label{eq:nmlongtimes}
n_m(t)\approx \left[1+\alpha K_{\rm mm}K_{\rm mc}\gamma^{-1}n_1^2(t-t_1)\right]^{-1/2},
\end{equation}
where $n_m(t_1)=n_1$ for some time $t_1>t'$ at which the long-times behavior has already been reached. $\alpha$ acts as an empirical  correction factor that accounts for \eq{eq:nmovernc} not reaching the dynamical equilibrium quite yet.

The time evolution of the molecular density is vastly influenced by
external electric fields, scaling as $d^{4L+2}$ for dipole moment $d$
and partial wave $L$ \cite{PhysRevA.84.062703}. As a result, for our
example in \fig{fig:moleculardensity}(a), the molecular density after
its initial, fast decay is almost cut in half for fields
$\gtrsim10$\,kV/cm. Even the field-free case is not free of losses due
to complex formation; however, over 90\% of the initial density is
retained. Hence, in spite of not being chemically reactive, ultracold RbCs
may still manifest substantial loss, which will become faster in an electric
field.

This is emphasized by \fig{fig:moleculardensity}(b), where we show the
ratio of total molecules bound in complexes to free molecules, for the
same set of field strengths as in panel (a). As expected from
\fig{fig:moleculardensity}(a), for zero electric field only a very small
fraction of the molecules is bound in complexes. This changes drastically,
however, once the field is turned on. For 10\,kV/cm, half the molecules
are trapped inside collision complexes at any given time.

\begin{figure}
\includegraphics[width=8.5cm]{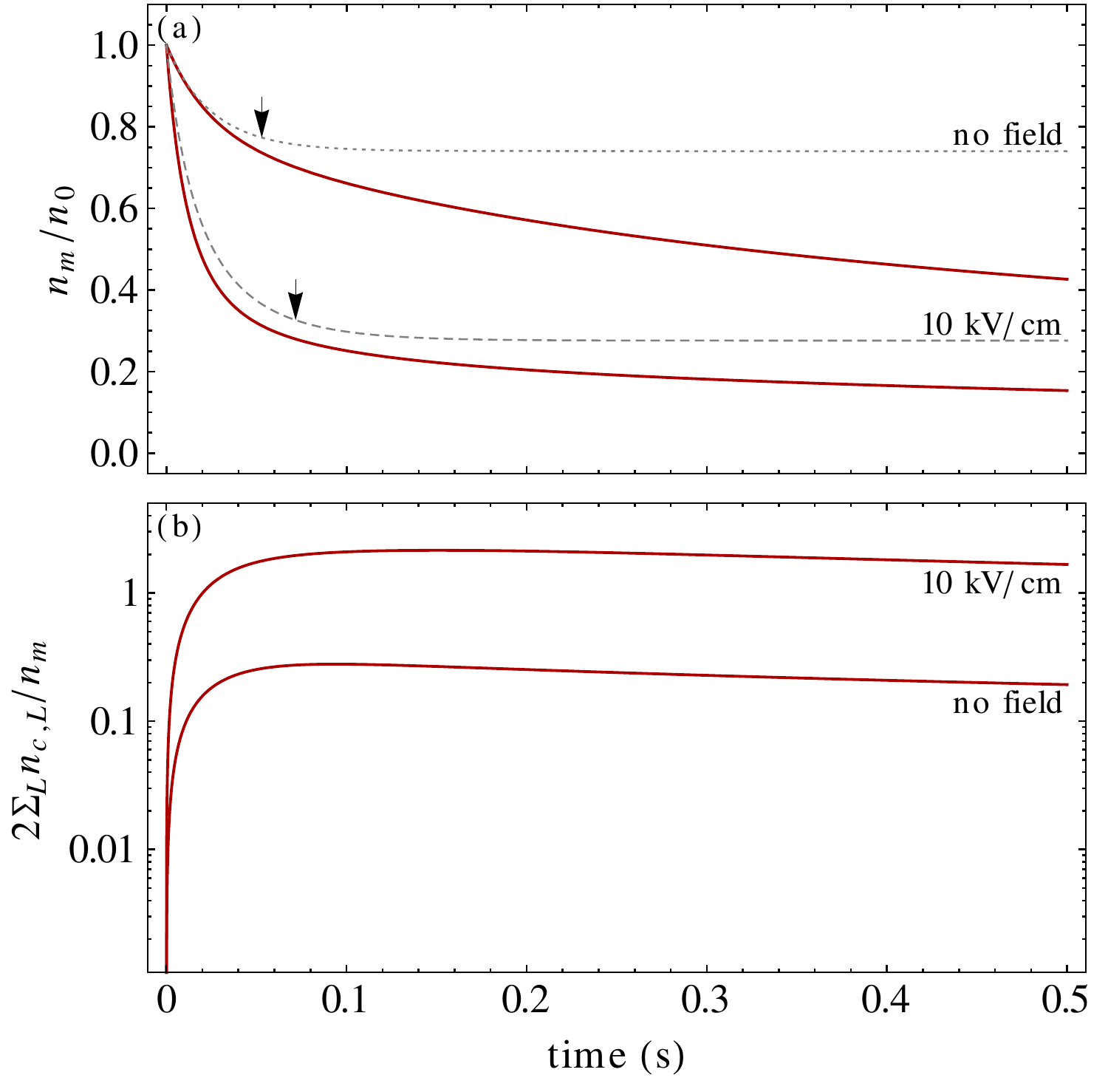}
\caption{Same as in \fig{fig:moleculardensity} but for a higher initial density of $n_0=10^{11}\mathrm{cm}^{-3}$.
\label{fig:moleculardensityhighn0}}
\end{figure}

The apparent loss of molecules depends on the magnitude of any applied electric field, but also on the initial density $n_0$ of molecules, cf.\ \eq{eq:nmfinal}. In \fig{fig:moleculardensityhighn0} we show the results for $n_0=10^{11}\mathrm{cm}^{-3}$, that is, increased by one order of magnitude compared to \fig{fig:moleculardensity}. Now, even zero electric field leads to significant molecule loss. For 10\,kV/cm, after only half a second fewer than 20\% of the initial molecular density is retained.

Experimental data on loss as in Figs.
(\ref{fig:moleculardensity},\ref{fig:moleculardensityhighn0})
can strongly constrain the parameters of the model
 At short times, the initial decay is given by $-K_{\rm mm}n_0$. With the initial density $n_0$ usually well known, one can therefore extract $K_{\rm mm}$ straightforwardly. At intermediate times, $n_m(t)$ turns from its initial drop to its long-time behavior. This timescale, \eq{eq:tprime}, is proportional to the lifetime of the complex. Hence, from experimental data, one could infer at least an estimate of the complex' lifetime. Finally, at long times the decay of the molecular density is well fitted by \eq{eq:nmlongtimes}, from which in turn the molecule-complex collision rate constant $K_{\rm mc}$ can be extracted.

\section{Influence on scattering of reactive molecules}

The situation is completely different for a species that is chemically reactive
at zero temperature, such as KRb.  In this case, transition state theory
dictates that the number of open channels $N_o$ includes also the product
channels.  These appear to be shockingly numerous, considering that
the exoergicity of ground state KRb collisions is only 10.4 $\text{cm}^{-1}$
\cite{PhysRevA.82.042707}.  Even within this small energy release, it is
possible, in principle, to produce K$_2$ molecules with rotation
quantum number up to $n_{K_2} = 13$, or Rb$_2$ molecules up to
$n_{Rb_2} = 20$, or
any energetically allowed combination. Moreover, the products can have
any reasonable partial wave angular momentum $L_{K_2 Rb_2}$ of the products
about each other, provided that the total angular momentum
${\bf J} = {\bf n}_{K_2} + {\bf n}_{Rb_2} + {\bf L}_{K_2 Rb_2}$
is conserved.

Again assuming ergodicity in all degrees of freedom, we obtain $N_o$ by
simply counting all possible exit channels consistent with
conservation of angular momentum and energy, constructing molecular levels from
the potentials in \cite{falke:012503,PhysRevA.82.052514,pashov:022511}.
The result is a vast number of possible channels, which grows rapidly as a
function of total angular momentum $J$.
Accounting for all these possibilities, the resulting number of
possible product channels are listed in Table \ref{tab:productchannels}
for various total $J$ but fixed $M=0$. No enhancement due to nuclear
spins is considered here.

Also shown in the table is the corresponding RRKM
decay rate into product channels.  This width far exceeds the mean level spacing
(by a factor of $N_o$, in fact), and renders the individual resonances
unobservable.  In fact, in this limit one expects collision cross sections
to exhibit Ericson fluctuations, on a scale comparable to $\Gamma$ itself.
Inasmuch this width is already of order $\mu$K (or $\sim 10$ Gauss
in magnetic field), it is unlikely that any structure will be seen at all.
Again, we are back to the simpler situation of studying non-resonant cross
sections.

Indeed, the occurrence of many exit channels implies that the decay rate of
the complex is extremely rapid, so rapid that the states of the complex
are not significantly populated at all.  An alternative way to view
this circumstance is to note that in the statistical theory
the probability of chemical reaction is $N_o/(N_o+1) \approx 1$, whereas the
probability of elastic scattering back to the single initial channel
is $1/(N_0 + 1) \ll 1$.  Thus the scattering leads to unit probability of
chemical reaction, as posited in Refs.\
 \cite{PhysRevA.84.062703,
Ospelkaus2010,Ni2010,PhysRevA.82.020703,PhysRevLett.104.113202,PhysRevA.81.022702,Miranda2011}. 
In fact, Ref.\ \cite{PhysRevLett.104.113202} provides a universal analytic expression of the inelastic rate constant for $p$-wave scattering with unit reaction loss probability, for identical fermionic molecules,
\begin{equation}\label{eq:klossin}
K^\mathrm{in,uni}_{L=1}=12\pi\frac{\hbar}{m_r}a_1(k\bar{a})^2\Delta.
\end{equation}
Here, $\bar{a}=2\pi R_\mathrm{vdW}/\Gamma(1/4)^2$ and $\bar{a}_1=\bar{a}\,\Gamma(1/4)^6/[12\pi\Gamma(3/4)]^2$ are length scales of $s$- and $p$-wave scattering from a pure $C_6$ potential.

\begin{table}
\caption{Number of KRb + KRb $\rightarrow$ K$_2$ + Rb$_2$ product channels along with the predicted RRKM width, $\Gamma_\mathrm{RRKM}=N_o/2\pi\rho$, of the resonances. In calculating the latter, we use the densities of states provided in Table \ref{tab:dos}.
\label{tab:productchannels}
}
\begin{ruledtabular}
\begin{tabular}{cccc}
 $J$ & number of channels & $\Gamma_\mathrm{RRKM}(\mu\mathrm{K})$ & $\Gamma_\mathrm{RRKM}(\mathrm{G})$\\
\hline
 0 &  45055  & 2.21 & 7.78 \\
 1 &  131239 & 2.15 & 7.27 \\
 2 &  213521 & 2.11 & 7.42 \\
 3 &  291901 & 2.06 & 6.97
\end{tabular}
\end{ruledtabular}
\end{table}

\begin{figure}
\includegraphics[width=8.5cm]{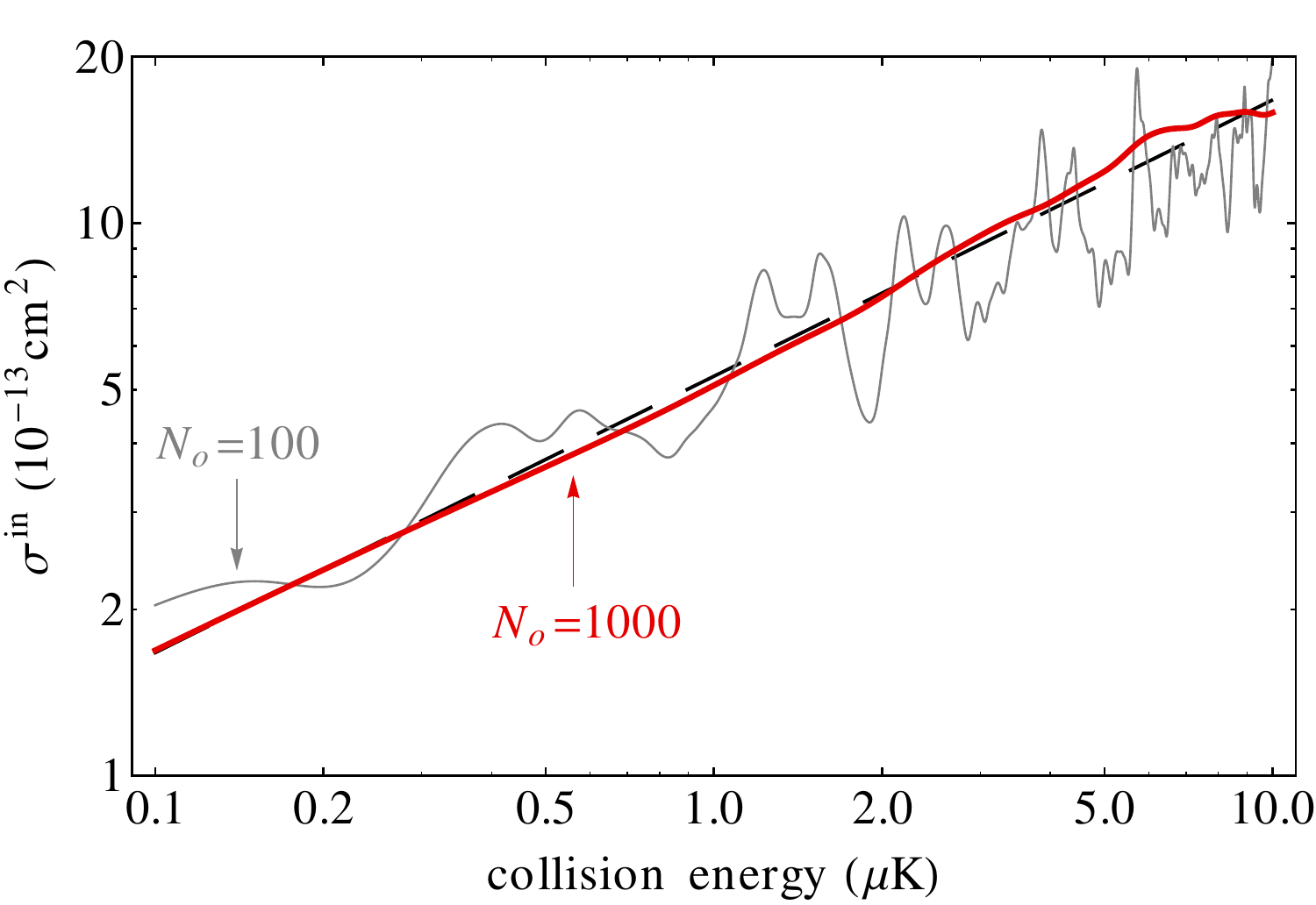}
\caption{(Color online) Inelastic $p$-wave scattering cross section mimicking ultracold KRb collisions with $N_o=100$ (light gray) and $N_o=1000$ [red (dark gray)] product channels. The dashed line represents the analytic prediction for unit loss probability, cf.\ \eq{eq:klossin}.
\label{fig:krbcsinel}}
\end{figure}

In \fig{fig:krbcsinel} we compare representative inelastic $p$-wave scattering
cross sections to the unit loss prediction \eq{eq:klossin}.
Rather than immediately employ the full number of open channels $N_o$ for
this case (which is technically challenging even within our simplified model),
we emphasize the trend for ever-larger $N_o$.  Thus the solid black line
shows an exemplary cross section for only $N_o=100$ open channels.
In this case Ericson fluctuations occur on a sub-$\mu$K scale and are
seen in the spectrum. However, even increasing the number of open channels
to $N_0 = 1000$ (red line) almost completely washes out these fluctuations.
Moreover, this result shows almost perfect agreement with the simple model
in Eqn. (\ref{eq:klossin}).  We find similar good agreement for different
realizations of the statistical spectrum.

We conclude from this result that the realistic $N_o$, which is much larger
still, will certainly lead to a featureless loss spectrum given by Eqn.
(\ref{eq:klossin}).  Thus the statistical model as deployed here
vindicates the models in Refs.
 \cite{PhysRevA.84.062703,
Ospelkaus2010,Ni2010,PhysRevA.82.020703,PhysRevLett.104.113202,PhysRevA.81.022702,Miranda2011}.

\section{Summary}

For collisions of alkali-metal dimer molecules, we have found that the
density of ro-vibrational states is enormous, far too large to probe
individual resonances
even at the sub-$\mu$Kelvin energy resolution afforded by ultracold
temperatures.  Because of this circumstance, resonant collision
rates are always averaged over many resonances,
and the theoretical description of scattering is greatly simplified.
Thus broad general conclusions can be drawn.
For the case of chemically reactive molecules, the formation of a
resonant state ensures that the atoms have ample opportunity to find their
way into the product channels, at least for reactions that are
assumed to be barrierless.  This in turn leads to essentially
unit probability of reaction in each collision event, consistent with
interpretations of recent experiments in ultracold KRb gases.

Strikingly, even  molecules that are {\it not} chemically reactive
at zero temperature, in the presence of this vast number of resonances,
behave {\it as if } they were chemically reactive, at least transiently.
These molecules are capable of sticking together for a finite
lifetime, which is dependent on the density of states.  The longer this
lifetime is, the more likely that the molecules bound in resonant
complexes will be struck by other molecules and lost.  Contrary to expectation,
it may therefore be necessary to shield even non-reactive molecules from
collisions by confining  them to 1D lattices and immersing them in
electric fields
\cite{Buchler07_PRL,Micheli10_PRL,Quemener10_PRA,Miranda2011,
Quemener11_PRA, Julienne11_PCCP}.

A main quantitative uncertainty in the results described here is whether
the full
density of ro-vibrational states is actually populated in a collision,
which may not be the case \cite{Nesbitt12_CR}.
If it is not, then the time during which the molecules are stuck together
reduces, and the loss rates may not be as great.
Thus measurements of loss may provide direct insight into the
four-body dynamics of the molecule-molecule complex.

\begin{acknowledgments}
The authors acknowledge financial support from the US Department of Energy and
the AFOSR. M.M.\ acknowledges financial support by a fellowship within the postdoctorate program of the German Academic Exchange Service (DAAD).
\end{acknowledgments}

\end{document}